# Radiomics and artificial intelligence analysis of CT data for the identification of prognostic features in multiple myeloma

Daniela Schenone[a], Rita Lai[b], Michele Cea[c], Federica Rossi[c], Lorenzo Torri[c], Bianca Bignotti[c,d], Giulia Succio[d], Stefano Gualco[c], Alessio Conte[c], Alida Dominietto[d], Anna Maria Massone[a,e], Michele Piana[a,e], Cristina Campi[f], Francesco Frassoni[a], Gianmario Sambuceti[c,d], Alberto Stefano Tagliafico[c,d]

[a]Dipartimento di Matematica, Università di Genova, Genova, Italy; [b]Dipartimento di Informatica, Bioingegneria, Robotica e Ingegneria dei Sistemi, Università di Genova, Genova, Italy; [c]Dipartimento di Scienze della Salute, Università di Genova, Genova, Italy; [d]IRCCS Ospedale Policlinico San Martino, Genova, Italy; [e]CNR - SPIN, Genova, Italy; [f]Dipartimento di Matematica "Tullio Levi Civita", Università di Padova, Padova, Italy

## ABSTRACT

Multiple Myeloma (MM) is a blood cancer implying bone marrow involvement, renal damages and osteolytic lesions. The skeleton involvement of MM is at the core of the present paper, exploiting radiomics and artificial intelligence to identify image-based biomarkers for MM. Preliminary results show that MM is associated to an extension of the intrabone volume for the whole body and that machine learning can identify CT image features mostly correlating with the disease evolution. This computational approach allows an automatic stratification of MM patients relying of these biomarkers and the formulation of a prognostic procedure for determining the disease follow-up.

## 1. INTRODUCTION

Multiple Myeloma (MM) is a blood cancer implying bone marrow involvement, renal damages and osteolytic lesions. Skeleton involvement of MM is at the core of the present paper, which introduces artificial intelligence and radiomics approaches for the identification of image-based biomarkers for MM. CT data already allows early identification of MM progression and the assessment of the positive response to chemotherapy. However, CT potentials in this field are far to be fully exploited; in particular, there is currently no image-based prognostic index predicting the evolvement of MM. We investigate whether this kind of indices can be identified in either the volumetric dimension of the intrabone space for the whole skeleton asset, or in the imaging features extracted from the focal lesion.

Image processing tools for the analysis of FDG-PET/CT data have been recently developed quantitatively assessing the composition of the skeleton asset and of FDG metabolism in bone marrow. These algorithms utilize pattern recognition in whole-body CT images of patients to segment the compact bone tissue from the bone marrow hosting intrabone volume. In this way, a normalcy database has been constructed [1], which has been utilized to identify the intra-bone volume as a prognostic marker for chronic lymphatic leukemia (CLL) [2] and to associate such alterations to functional modifications in the bone marrow asset of allogeneic transplant [3]. Even more recently, machine learning methods have been applied to features extracted from different imaging modalities within the framework of radiomics paradigms for disease assessment. A few of these methods perform the evaluation of bone formation, regeneration, and asset and, in some cases, these algorithms are able to identify image features that mostly impact the prediction of the disease follow-up. However, none of these methods have been applied so far to CT data from MM patients and therefore the purpose of this paper is two-fold:

- To verify that, as in the case of CLL, intrabone volume represents a reliable biomarker for determining the evolution of MM.
- To verify that the image properties of the compact bone focal lesions can be used for patients' stratification.

The plan of the paper is as follows. Section 2 illustrates the data used for the analysis and the computational methods for their processing. Section 3 describes the results obtained from the data analysis. Section 4 provides some comments about these results. Our conclusions are offered in Section 5.

Table 1. Minimal and standard Computed Tomography technical parameters for inclusion

| Number of detector rows* | 16 or more up to 128 |
| --- | --- |
| Minimum Scan coverage* | Skull base to femur |
| Tube voltage(kV)/time-current product (mAs) | 120/50–70, adjusted as clinically needed |
| Reconstruction convolution kernel | Sharp, high-frequency (bone) and smooth (soft tissue). Middle-frequency kernel for all images are adjusted by the radiologist as deemed necessary |
| Iterative reconstruction algorithms | Yes (to reduce image noise and streak artifacts) |
| Thickness* | ≤5 mm |
| Multiplanar Reconstructions (MPRs) | Yes (sagittal, coronal and parallel to long axis of proximal limbs) |
| Matrix, Rotation time, table speed, pith index | 128x128, 0.5 s, 24mm per gantry rotation, 0.8 |

## 2. MATERIALS AND METHODS

### 2.1 Study design, inclusion criteria and CT imaging

The study was performed in accordance with the current version of the Declaration of Helsinki and the International Conference on Harmonization of Good Clinical Practice Guidelines. All patients signed a written informed consent form, encompassing the use of anonimized data for retrospective research purposes, before CT examination. Radiomic analysis was applied to CT data collected in the clinical workup and did not influence patient care in any way.

As far as the design of this retrospective study and the corresponding inclusion criteria are concerned, we have considered 25 consecutive patients (mean age, 62 years ± 8.3; range, 35–70 years) admitted to the IRCCS Policlinico San Martino Hospital because they were suspected of having MM in the last five years. Inclusion criteria were baseline whole-body CT available and retrievable from the Hospital PACS or available from outpatient clinic. The imaging technical standards are minimal and reported in Table 1.

Data obtained in patients' population were compared to data from 102 control subjects with no history of hematological disease, selected from a previously published normalcy database [1].

## 2.2 Image analysis: whole body data

First, the global CT information was processed in order to determine the intraosseous volume potentially available for bone marrow. In order to determine this intrabone volume (IBV) we utilized an already published software code whose underlying assumption is that the Hounsfield value is highest in compact bone among all tissues. Given the fact that attenuation coefficients differ in bones belonging to different districts, we could not apply a mere thresholding technique and therefore we relied on a pattern recognition process based on active contours (see Figure 1). More precisely, the process starts with the unique human intervention asking the operator to draw a loose region of interest around the skull vertex as a starting reference. Then, the functional representing the energy of a curve surrounding the bone profile is iteratively minimized thus determining a sequence of active contours that progressively adapt themselves onto the compact bone border. This procedure is automatically replicated to all slices. In particular, the optimized active contour obtained at the end of the processing of the first slice is utilized as initialization contour for the successive slice. The final 3D result is thus displayed to the operator for the removal of not-bone calcified regions as well as for the check of the appropriate recognition of the spinal canal as extraosseous space.

The output of the software is therefore the quantitative assessment and the 3D representation of three different volumes: 1) the whole skeleton; 2) the compact bone tissue and 3) the space potentially available for BM. The application of these procedures to all slices of acquisition permits to evaluate the whole skeleton and each slice to exclude non-bone calcified regions or possible inclusion of spinal canal.

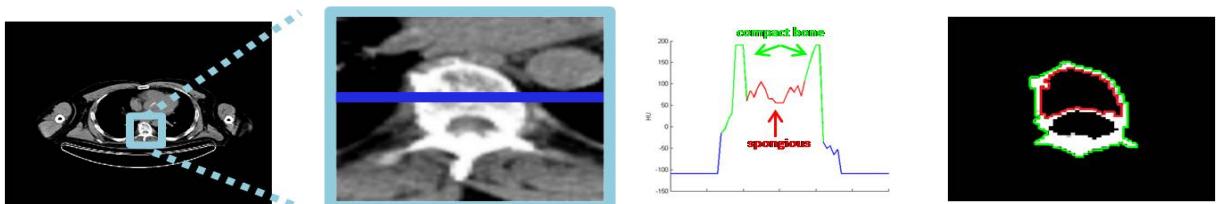

Figure 1. Pictorial description of the pattern recognition approach for IBV identification.

## 2.3 Image analysis: radiomics of focal lesion

For each MM patient, and utilizing an open source software tool for radiomics (https://www.radiomics.io/slicerradiomics.html), we have also applied pattern recognition algorithms on several primary lesions present on the compact bone tissue in order to extract image properties that can be used as input features for machine learning algorithms (see Figure 2). Specifically, for each focal lesion radiomics extracts 140 features that may be given as input to either supervised or unsupervised artificial intelligence algorithms for automatic stratification.

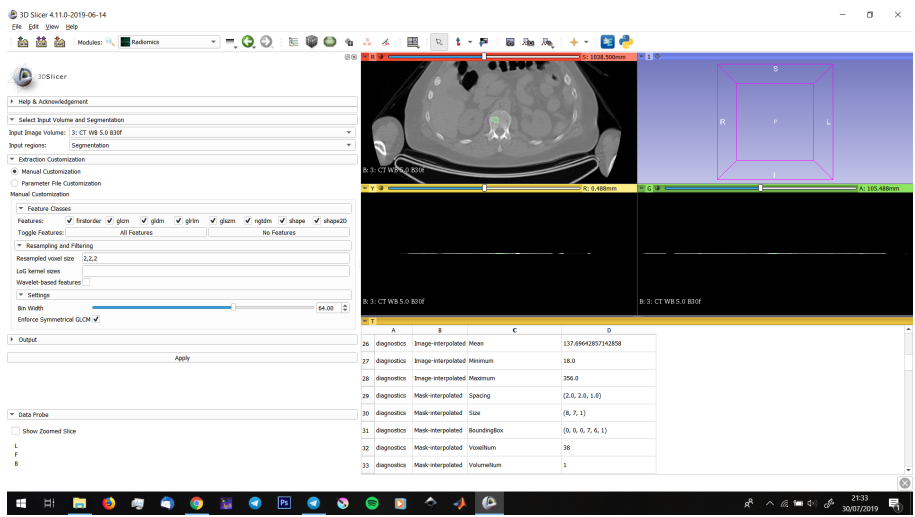

Figure 2. Outcome of the radiomics tool in the case of a focal lesion in the vertebral body.

## 2.4 Post-processing: patients' stratification

The post-processing relied on a standard unsupervised clustering analysis of both properties provided by the software tool realizing the assessment of the skeleton asset and features provided by radiomics of the focal lesions. To this aim we applied a hard C-means algorithm in which the number of classes is fixed a priori and each data sample may belong to just one cluster. Extension of this approach may include fuzzy clustering and possibilistic clustering, in which the two main constraints are relaxed, i.e. the number of classes adapts itself to the data properties and each sample may belong to different classes with a specific probability.

## 3. RESULTS

For each MM patient, the software for the quantitative assessment of the whole skeleton asset computed the following parameters:

- The intrabone volume (the volume at disposal of the spongiosa) normalized with respect to the ideal body weight (N-IBV)
- The volume occupied by the compact bone normalized with respect to the ideal body weight (N-CBV).
- The whole skeleton volume normalized with respect to the ideal body weight (N-BV = N-IBV + N-CBV).
- The fraction of volume occupied by trabecular bone with respect to the overall skeleton volume (%IBV).
- The average Hounsfield value for the compact bone and its standard deviation.
- The average Hounsfield value for the spongiosa and its standard deviation.

Trabecular bone was expanded in MM patients and occupied a larger fraction of the whole skeleton asset with respect to the control subjects whose data are contained in a normalcy database made of 102 control subjects (see Figure 3). We obtained N-IBV = 31 ± 5 mL/Kg (vs N-IBV = 27 ± 8 mL/kg), N-CBV = 51 ± 8 mL/kg (with respect to 59 ± 9 mL/kg), and N-BV = 81 ± 12 mL/kg (with respect to 86 ± 12 mL/kg). Further, in MM patients this trabecular bone occupies a larger fraction of the skeleton with respect to controls (%IBV = 38 ± 2 vs %IBV = 31 ± 7).

*The application of the hard C-means clustering to some of the properties extracted from the whole skeleton asset.* Table 1 illustrates the results of this stratification process: on the basis of five of these properties the 25 patients were partitioned into 2 classes made of 15 (cluster 1) and 10 patients (cluster 2), respectively. Interestingly, an a posteriori check of the populations of these two clusters showed that all four patients that underwent relapse of the disease within the three months following the diagnosis (see Figure 4). This is preliminary result is not confirmed by an analogous analysis made on features extracted by radiomics from ROIs including the focal lesions. Figure 5 shows that the partition obtained in the case of these local features is not coherent with the one obtained in the case of features corresponding to global properties associated to the whole skeleton asset. The histograms in the figure have been realized by means of 100 runs of the hard C-means code corresponding to 100 initialization of the clusters' centroids and reporting how many times the samples is classified in each one of the two clusters.

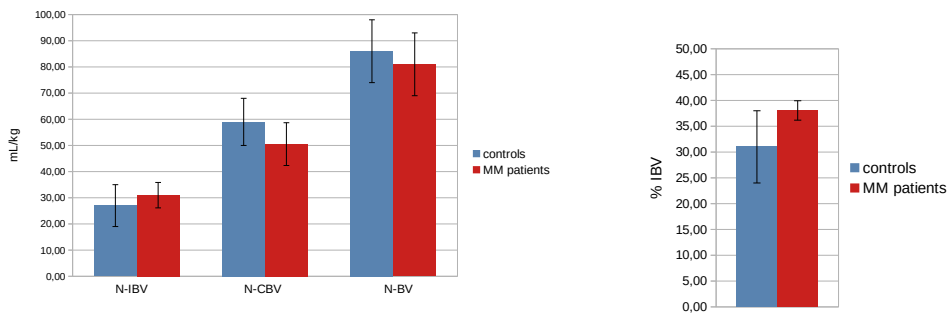

Figure 3. Quantitative assessment of the whole bone asset in its different components and comparison between control and MM subjects. Left panel: normalized intrabone volume (IBV), normalized compact bone volume (CBV) and normalized total bone volume (BV). Right panel: rate of IBV with respect to BV. All values have been averaged over the corresponding population and the standard deviation has been computed.

## 4. COMMENTS AND CONCLUSIONS

A standard Student t-test shows that the statistical significance of the differences between the two populations is higher for N-CBV and for %IBV ($p<0.05$) with respect to N-IBV ($p=0.05$). However, this test represents a first hint supporting the prognostic value of these three global parameters. The global properties associated to the overall compact bone and trabecular tissues are also effective to stratify the patents' population. Indeed, the results of the application of hard C-means on 25 sets made of five over the overall 5 global features, show that all four patients who underwent MM relapse are classified in the same cluster. Interestingly enough, this stratification power seems not to be shared by the 26 radiomics features extracted from the focal lesion associated to each patient: clearly, the histograms in Figure 5 are not coherent with the class identified in Figure 4.

Table 2. Results of the clustering process applied against five features extracted by means of the software tool for the assessment of the whole skeleton asset. Specifically, f1 is the standard deviation associated to the average Hounsfield value for the trabecular bone; f2 is the average volume expressed in $cm^3$; f3 is the average Hounsfield value for the compact bone; f4 is the standard deviation associated to the average Hounsfield value for the compact bone; f5 is the fraction of volume occupied by trabecular bone with respect to the overall skeleton volume; C-means is the identification of cluster analysis (cluster 1 or cluster 2); Relapse 0/1 indicate if the patient underwent MM relapse.

| patient | f1 | f2 | f3 | f4 | f5 | C-means | relapse 0/1 |
|---|---|---|---|---|---|---|---|
| 1 | 152,6 | 1547 | 494,6 | 364,6 | 37,3 | 1 | 0 |
| 2 | 156,8 | 1748 | 515,0 | 379,6 | 38,5 | 1 | 0 |
| 3 | 161,9 | 1401 | 483,6 | 430,5 | 39,8 | 1 | 0 |
| 4 | 174,5 | 1978 | 641,5 | 437,6 | 39,8 | 2 | 0 |
| 5 | 165,2 | 1181 | 555,9 | 405,6 | 36,6 | 1 | 0 |
| 6 | 166,0 | 2889 | 606,5 | 453,2 | 44,5 | 2 | 0 |
| 7 | 187,3 | 2029 | 633,8 | 448,2 | 38,3 | 2 | 1 |
| 8 | 173,1 | 2340 | 582,4 | 436,2 | 39,5 | 2 | 1 |
| 9 | 160,6 | 2558 | 563,5 | 395,4 | 39,3 | 1 | 0 |
| 10 | 163,9 | 1473 | 571,5 | 442,1 | 35,6 | 1 | 0 |
| 11 | 153,6 | 1107 | 514,7 | 376,2 | 37,7 | 1 | 0 |
| 12 | 170,2 | 2822 | 531,1 | 420,0 | 37,5 | 1 | 0 |
| 13 | 165,3 | 1674 | 637,5 | 407,6 | 37,8 | 2 | 0 |
| 14 | 154,1 | 1843 | 481,0 | 354,8 | 36,1 | 1 | 0 |
| 15 | 189,4 | 2534 | 685,4 | 428,3 | 35,2 | 2 | 1 |
| 16 | 162,5 | 1285 | 579,2 | 401,0 | 37,2 | 1 | 0 |
| 17 | 209,5 | 1282 | 610,9 | 607,6 | 37,3 | 2 | 0 |
| 18 | 165,7 | 1451 | 573,6 | 419,9 | 38,4 | 1 | 0 |
| 19 | 172,4 | 2325 | 582,1 | 406,0 | 39,9 | 2 | 0 |
| 20 | 171,0 | 1931 | 624,3 | 456,5 | 35,2 | 2 | 0 |
| 21 | 163,9 | 2292 | 565,6 | 423,8 | 37,6 | 1 | 0 |
| 22 | 146,0 | 1450 | 504,1 | 369,3 | 38,0 | 1 | 0 |
| 23 | 173,0 | 1649 | 577,9 | 410,6 | 35,4 | 1 | 0 |
| 24 | 167,0 | 1764 | 594,9 | 436,8 | 38,9 | 2 | 1 |
| 25 | 148,0 | 2241 | 522,9 | 393,8 | 38,1 | 1 | 0 |

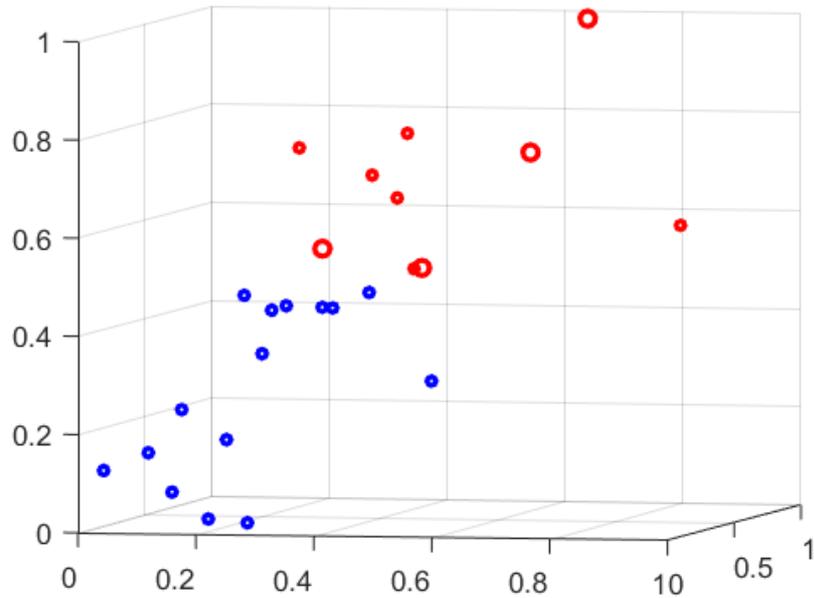

Figure 4. Patients' stratification obtained by applying a hard C-means clustering code on the five features described in Figure 3 and associated to the 25 MM subjects (the representation involves just three features). Blue small circle: patients classified in cluster 1; red small circle: patients classified in cluster 2; red big circle: patients with relapsed MM within three months after diagnosis.

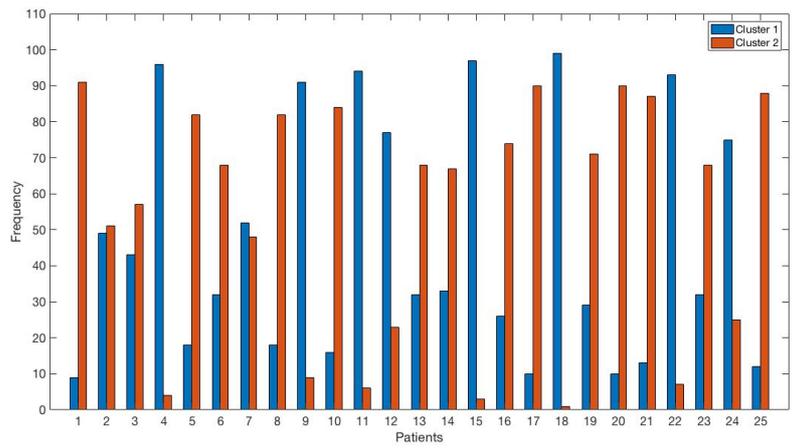

Figure 5. Result of the application of the hard C-means algorithm on the 26 features extracted by an open source code for pattern recognition from CT images of the focal lesion. In the figure we have highlighted the four patients who actually underwent MM relapse.

## REFERENCES

[1] Sambuceti, G., Brignone, M., Marini, C., Massollo, M., Fiz, F., Morbelli, S., Buschiazzo, A., Campi, C., Piva, R., Massone, A. M., Piana, M. and Frassoni, F. "Estimating the whole bone marrow asset in humans by a computational approach to integrated PET/CT imaging", Eur. J. Nucl. Med. Mol. Imag., 39(8) 1326-1338 (2012).

[2] Fiz, F., Marini, C., Piva, R., Miglino, M., Massollo, M., Bongioanni, F., Morbelli, S., Bottoni, G., Campi, C., Bacigalupo, A., Bruzzi, P., Frassoni, F., Piana, M. and Sambuceti, G., "Adult advanced chronic lymphocytic leukemia: computational analysis of whole-body CT documents a bone structure alterantion", Radiology, 271(3) 805-813 (2014).

[3] Fiz, F., Marini, C., Campi, C., Massone, A. M., Podestà, M., Bottoni, G., Piva, R., Bongioanni, F., Bacigalupo, A., Piana, M., Sambuceti, G. and Frassoni, F., "Allogeneic cell transplant expands bone marrow distribution by colonizing previously abandoned areas: an FDG PET/CT analysis", Blood, 125(26) 4095-4102 (2015).